\documentclass[aps,prc,showpacs,twocolumn]{revtex4}
\usepackage{graphicx}
\usepackage{dcolumn} 
\usepackage{bm}
\usepackage{hyperref}
\usepackage{natbib}
\usepackage{amsmath}
\usepackage{multirow}
\usepackage{color}
\begin{document}
\title{$0\nu\beta\beta$ to the first $2^+$ state with two-nucleon mechanism for L-R symmetric model}
\author{Dong-Liang Fang$^{a,b}$ and Amand Faessler$^{c}$} 
\address{$^a$Institute of Modern Physics, Chinese Academy of Science, Lanzhou, 730000, China}
\address{$^b$University of Chinese Academy of Sciences, Beijing, 100049,China}
\address{$^c$Institute for theoretical physics, T\"ubingen University, D-72076, Germany}
\begin{abstract}
We develop the formalism for calculating the decay rate of neutrinoless double beta decay to the $2^+$ excited states within L-R symmetric model. 
We consider the effects from induced hadronic currents up to NLO. The QRPA method in a spherical basis is adopted for the nuclear many-body calculation and the corresponding nuclear matrix elements are given. Also, the phase space factors are obtained with numerical electron wave functions. Our results suggest that the nuclear matrix elements are nucleus dependent and they are generally smaller than that of the decay to the ground states. 
And finally, we give a naive analysis of how current experiment data constrains the L-R symmetric model. 
\end{abstract}
\maketitle

\section{Introduction}
New physics beyond the standard model is always the hottest topic in particle physics. And the origin of the neutrino mass could be the one among the most important questions in this area. Perhaps, the most promising explanation for a small neutrino mass is the see-saw mechanism. The see-saw mechanism can usually be divided into different categories. To realize such mechanisms, different new physics models are proposed. One of the appealing proposals is the L-R symmetric model \ \cite{MS81} with the underlying gauge symmetry SU(2)$_L\times$SU(2)$_R\times$U(1)$_{B-L}$. With two more Higgs particles introduced, one can naturally incorporate the seesaw mechanism into this model. There are different phenomenologies related to this model. One of the important consequences is the existence of the so-called neutrinoless double beta decay ($0\nu\beta\beta$-decay). 

However, the underlying mechanism of $0\nu\beta\beta$-decay may not be unique.
Thus, it is important to find a way to identify the decay mechanisms by future measurements. There are several proposals for this purpose, such as comparing the ratios of the decay rates for different 
candidates \ \cite{LRS15} or measuring the spectra and the angular correlations of the emitted electrons\ \cite{GLS22}. As an alternative, one also suggests to compare the decay rates of decays to ground and excited states\ \cite{DLZ11}, especially between the ground states and the $2^+$ excited states \ \cite{GLS22}. 

In minimal L-R symmetric model, for the decay to ground state, due to neutrino propagator's helicity suppression by the mass mechanism, the non-helicity-suppressed $q$-term may play a dominant role\ \cite{DKT85}. Under such scenario, the emission of P-wave electrons will surely lead to visible effect on the angular correlation of the double $\beta$ spectrum, this is investigated in \ \cite{SDS15}. Also,  for another decay mode -- neutrinoless double beta decay to the $2^+$ excited states (Hereafter $0\nu\beta\beta(2^+)$), contributions from such P-wave electrons will become dominant. While different to the decay to ground states, the helicity suppressed $m_e$ terms are negligible since these terms come from the NLO parts of the hadronic current. So with such new physics model, we may probe the underlying mechanism by comparing these different decay modes. Unfortunately, $0\nu\beta\beta\beta(2^+)$ is actually rarely investigated, and the only calculation available is done with the Projected HFB approach \ \cite{Tom88}. PHFB calculation suggests that the decay is highly suppressed since the NME is several orders of magnitude smaller than that of the decay to ground states (Hereafter $0\nu\beta\beta(0^+)$). This smallness is caused partly by the suppression from each part of the NME and partly by the cancellation among them. On the other hand, in\ \cite{Tom99}, one finds that although the neutrino mass mechanism can contribute to this decay mode, their NME is about two order of magnitude smaller than the $q$ mechanism. These together make it impossible to observe $0\nu\beta\beta(2^+)$. 

Nevertheless, our recent calculations \cite{FF21} suggest that the NME in \cite{Tom88} is underestimated, our results are orders of magnitude larger than that of \cite{Tom88} and cancellations among different components are not observed especially for $M_{\eta'}$. These new results may suggest that decay to the first $2^+$ states is not that heavily suppressed as previously expected. In all previous calculations, only the vector and axial-vector parts of the hadronic current are considered. Another important component, namely the pseudo-scalar part from the pion pole\ \cite{GT58} is not taken into account. As suggested in \ \cite{CDV17}, this pseudo-scalar piece is accounted as a LO contribution like the vector and axial-vector parts. In most $0\nu\beta\beta$ calculations, the NLO weak-magnetism contribution is also taken into account. In this work, we incorporate all these parts into the calculation and study their effects on the NME's.

The calculation of NME relies on various nuclear many-body approaches, and we limit our discussions to traditional ones with effective nuclear forces, leaving out those of {\it ab initio} methods using a nuclear force starting from bare nucleon-nucleon interaction. For the decay to the ground states with the standard neutrino mass mechanism, several calculations have been done over the decades. First we mention the time-consuming large scale shell model (LSSM) calculations \ \cite{CNP99,Men17}, which takes advantage of the existence of the shell gap, and separates  the particles of the core and the nucleons of the valence part. 

Apart from LSSM calculations, which are applicable to limited cases due to the large computation requirement, several other  methods can be applied to more occassions, {\it e.g.} the IBM-2 method \ \cite{BKI13}, the DFT methods of non-relativistic \ \cite{RM10} and relativistic versions \ \cite{SYR14}, the project HFB method \ \cite{RCC13}, as well as the QRPA method \ \cite{SPV99,SRF13,HS15,FFS18}, which takes intermediate states into account.
For recent reviews of these calculations, we refer to \ \cite{YMN21,ABD22}. 
 
Meanwhile, the NME calculations for the non-standard LR-symmetric model with the inclusion of the non-helicity-suppressed $q$ terms are less frequently considered for $0\nu\beta\beta(0^+)$. Recent calculations have been done by QRPA \ \cite{MBK89,SDS15,SSD17}, LSSM \ \cite{HN15,HN18,AH19,SIR20,IS21} as well as by PHFB \ \cite{RCC19}. Especially for \ \cite{CNP96,HN15,HN18,AH19}, contributions from SM effective field theory has been thoroughly analyzed besides the LR-symmetric model. There are also other calculations starting from SM effective field theories which have different expressions as traditional LR symmetric models\cite{CDV17, CDV18}.

Compared to above calculations of decay to the ground states, the decay to the excited $2^+$ state is rarely discussed, and the most recent calculations are about two decades ago \ \cite{Tom99} as mentioned above. To investigate this special decay mode, we adopt the QRPA method with realistic nuclear forces \ \cite{FF21}. Our previous results suggest that the final NMEs are larger than expected with only the vector and axial vector parts of the hadronic currents included. In this work, we include more components up to NLO to make a more thorough investigation.

This article is arranged as follows: At first we present the formalism we use. It is followed by the results for the  phase space factors and the nuclear matrix elements. Then we discuss constraints  on the L-R model parameters from current results. Finally we give a conclusion as well as an outlook.

\section{Formalism}
In the L-R symmetric models, such as SU(2)$_L\times$SU(2)$_R\times$U(1)$_{B-L}$, after two successive spontaneous symmetry breakings, the left- and right- handed gauge bosons acquire masses through the Higgs mechanism, and in general the left- and right- handed gauge bosons are mixed  \ \cite{MS81,DKT85}:
\begin{eqnarray}
\left( \begin{array}{c}
W_L \\ 
W_R
\end{array}\right)
= \left( \begin{array}{cc}
\cos \xi & -\sin \xi \\ 
\sin \xi & \cos \xi
\end{array}\right)
\left( \begin{array}{c}
W_1 \\ 
W_2
\end{array}\right)
\end{eqnarray}
Here $\xi$ is the mixing angle and $W_1$, $W_2$ the mass eigenstates of W bosons.

The neutrinos acquire masses through their Yukawa coupling with Higgs bosons \ \cite{MS81}:
\begin{eqnarray}
\left( \begin{array}{c}
\nu_L \\ 
N_R
\end{array}\right)= \left( \begin{array}{cc}
U & U' \\ 
V' & V
\end{array}\right)
\left( \begin{array}{c}
\nu_M \\ 
N_M
\end{array}\right)
\end{eqnarray}
Where $\nu_L^T=(\nu_e, \nu_\mu, \nu_\tau)$ and $N_R^T=(N_e, N_\mu, N_\tau)$ are the three flavor left- and right- handed neutrinos. $\nu_M$ and $N_M$ are their light and heavy mass eigenstates. The see-saw mechanism can be naturally fulfilled in this model \ \cite{MS81}.

Starting from the left- and right- gauge-fermion interactions, the effective weak Hamiltonian can be written following the definition in \ \cite{DKT85}:
\begin{eqnarray}
H_{\rm eff}=\sqrt{\frac{1}{2}} G_F \cos\theta_C (j_{L\mu} \tilde{J}^{\mu}_L+ j_{R\mu} \tilde{J}^{\mu}_R) +h.c.
\end{eqnarray}

Where the lepton currents are:
\begin{eqnarray}
j_{L(R)}^\mu(\vec{x})=\bar{\psi}_e(\vec{x}) \gamma^\mu P_{L(R)} \psi_{\nu} (\vec{x})
\end{eqnarray}
With $P_L=(1-\gamma_5)/2$ and $P_R=(1+\gamma_5)/2$ respectively.


In current model, the hadronic currents have the form:
\begin{eqnarray}
\tilde{J}_{L\mu}&\approx&J_{L\mu} \nonumber \\
\tilde{J}_{R\mu}&\approx&\eta J_{L\mu}+\lambda J_{R\mu}
\end{eqnarray}
Here 
\begin{eqnarray}
\eta\equiv - (g_R/g_L)\tan\xi[1-(M_1/M_2)^2]/\\ \nonumber [1+\tan^2\xi(M_1/M_2)^2]  \hspace{1.0 cm}  \\
\lambda\equiv (g_R/g_L)^2[(M_1/M_2)^2 +\tan^2 \xi]/\\ \nonumber [1+\tan^2\xi (M_1/M_2)^2] \hspace{1.0cm}\\ 
\nonumber
\end{eqnarray} 
Here $M_1$ and $M_2$ are the mass eigenvalues of $W_1$ and $W_2$ gauge bosons respectively.

Within the non-relativistic impulse approximation, under the Breit frame, the left- or right- handed hadronic currents have the form: 
\begin{eqnarray}
J_{L\mu}(\vec{x})&=&(J_0(\vec{x}),\vec{J}_L(\vec{x}))
\nonumber \\
J_{R\mu}(\vec{x})&=&(J_0(\vec{x}),\vec{J}_R(\vec{x}))
\end{eqnarray}
Where
\begin{eqnarray}
J_0(\vec{x})&=&\sum_{n=1}^A g_V(q^2) \delta(\vec{x}-\vec{r}_n) \nonumber\\
\vec{J}_L&=&\sum_{n=1}^A -[g_A(q^2) \vec{\sigma}_n - g_P(q^2) (\vec{\sigma}_n\cdot\vec{q}) \vec{q} \nonumber \\
&+& i\frac{g_M(q^2)}{2m_p} (\vec{\sigma}_n\times \vec{q})]\delta(\vec{x}-\vec{r}_n) \nonumber \\
\vec{J}_R&=&\sum_{n=1}^A  [g_A(q^2) \vec{\sigma}_n - g_P(q^2) (\vec{\sigma}_n\cdot\vec{q}) \vec{q} \nonumber \\ 
&-& i\frac{ g_M(q^2) }{2m_p}(\vec{\sigma}_n\times \vec{q})]\delta(\vec{x}-\vec{r}_n)
\end{eqnarray}

According to angular momentum conservation, for $0\nu\beta\beta(2^+)$, the emitted electrons must be coupled to total angular momentum $J=2$, this suggests that the dominant contribution comes from the combination of decomposed partial waves $s_{1/2}$-$p_{3/2}$\ \cite{DKT85}. Substituting the hadronic currents into the S-matrix, the decay width can then be written as \ \cite{DKT85,Tom88}:
\begin{eqnarray}
\Gamma=G_1 |M_\lambda \langle \lambda \rangle -M_\eta \langle \eta \rangle |^2 + G_2 |M'_{\eta} \langle \eta \rangle|^2
\label{dw}
\end{eqnarray}
Here G's are phase space factors and M's are the NMEs. $\langle \lambda \rangle$ and $\langle \eta \rangle$ are new physics parameters connected to $\lambda$ and $\eta$ defined above as\cite{SDS15}: $\langle \eta \rangle=\eta |\sum_{j}U_{ej} V'^*_{ej}|$ and $\langle \lambda \rangle=\lambda |\sum_{j}U_{ej} V'^*_{ej} (g'_V/g_V)|$, here $g_V$ and $g'_V$ are the vector coupling constant for left- and right-handed currents respectively. U and V' are neutrino mass mixing matrix elements. {And j sums over the light neutrino mass eigenstates.}

The phase space factor(PSF) can be expressed as \ \cite{Tom88}:
\begin{eqnarray}
G_i&=&\frac{4 \pi}{\ln 2 ~ R_{\rm n}^4 } \int dE_1 dE_2 \frac{(G \cos \theta_C)^4}{32 \pi^6} f_i p_1 p_2 E_1 E_2 \nonumber \\
&\times& \delta(E_1+E_2-2 m_e - Q_{\beta\beta}(2^+))
\end{eqnarray}
Here
\begin{eqnarray}
f_1= 3 [|f^{-2-1}|^2+|f_{21}|^2+|f^{-1-2}|^2+|f_{12}|^2]\nonumber \\
f_2= 3 [|f^{-2}{}_{1}|^2+|f^{-1}{}_2|^2 + |f^{-2}{}_{1}|^2+|f^{-1}{}_{2}|^2]
\end{eqnarray}
For these $f$ functions such as $f_{12}$, we follow the convention in \ \cite{KI12}. In our definition, the phase space factor are with the unit of $y^{-1}$, this is obtained by dividing $R_{\rm n}^2$ to eq.(11) of \cite{Tom88}. Here $R_{\rm n}=1.2 A^{1/3} $fm is the conventionally defined nuclear radius. By deriving this, the no Finite de Broglie wave length correction (no FBWC) approximation is used \ \cite{DKT85}.

The expressions for the NME are much more complicated, and we follow the conventions in \cite{Tom88}, divide the NME into seven parts:
\begin{eqnarray}
M_\lambda=\sum_{i=1}^5 C_{\lambda i} M_{ i} \nonumber \\
M_\eta=\sum_{i=1}^5 C_{\eta i} M_{ i} \nonumber \\
M'_\eta=\sum_{i=6}^7 C'_{\eta i} M_{ i} \nonumber
\end{eqnarray}
For the coefficients $C$'s, we follow the definition of \cite{Tom88} too and they are tabulated in Table.\ref{coef}, where we have absorbed $g_A$ into the NME unlike the conventional treatment where coupling constants $g$'s are included in PSFs.

The NME can be further expressed as $M_i=\langle 2_f^+ || \mathcal{M}_i||0_i^+ \rangle$, and these operators $ \mathcal{M}_i$ can be expressed in a general form:
\begin{eqnarray}
\mathcal{M}_{i} = \frac{2R_{\rm n}}{\pi}\int \frac{q dq}{q+E_N} h_{i}(q,r) \mathcal{O}_i
\end{eqnarray}
Here $E_N=E_{mx}+M_m - (M_i +M_f+E_{2^+})/2$ is the intermediate state excitation energy relative to the initial and final states. And $h$ is usually called the neutrino potential and $ \mathcal{O}$ is the angular transition operator.

Where for different NME components, the detailed forms for neutrino potential $h_{i}$ is as follows:
\begin{widetext}
\begin{eqnarray}
h_1(q,r)&=&\frac{1}{g_A^2(0)}j_1(qr) [ g_{A}^2(q^{2}) + 2 \frac{g_A(q^{2}) g_P(q^{2}) q^2}{2m_p} -  \frac{ g_P^2(q^{2}) q^4}{(2m_p)^2}   - 2 \frac{g^2_M(q^{2}) q^2}{(2m_p)^2}) ] \nonumber \\
h_2(q,r)  &=&\frac{1}{g_A^2(0)} j_1(q r) [ g_A^2(q^2) - \frac{g_A(q^{2}) g_P(q^{2}) q^2}{2m_p} + \frac{1}{5}\frac{ g_P^2(q^{2}) q^4}{(2m_p)^2} +\frac{1}{5} \frac{g^2_M(q^{2}) q^2}{(2m_p)^2} ] \nonumber \\
h_3(q,r)  &=& \frac{1}{g_A^2(0)}\{ j_1(qr) [g_A^2(q^2)-\frac{g_{P}(q^2)g_{A}(q^2) q^2}{2m_p} +  \frac{1}{5} \frac{ g_P^2(q^{2}) q^4}{(2m_p)^2}+\frac{1}{5} \frac{g^2_M(q^{2}) q^2}{(2m_p)^2}]  
- j_3(qr) [ \frac{3}{35} \frac{ g_P^2(q^{2}) q^4}{(2m_p)^2} + \frac{3}{35} \frac{g^2_M(q^{2}) q^2}{(2m_p)^2} ] \} \nonumber \\
h'_3(q,r) &=&\frac{1}{g_A^2(0)} j_3(qr) [\sqrt{\frac{3}{5} } \frac{ g_P^2(q^{2}) q^4}{(2m_p)^2} + \sqrt{\frac{3}{5}} \frac{g^2_M(q^{2}) q^2}{(2m_p)^2} ] \nonumber \\
h_4(q,r) &=&\frac{g_V^2(q^2)}{g_V^2(0)} j_1(qr)  ,\quad 
h_5(q,r) =\frac{g_V(q^2) g_A(q^2)}{g_V(0)g_A(0)} j_1(qr)  , \quad
h_6(q,r) = h_7(q,r)= \frac{g_V(q^2) g_A(q^2) }{g_A^2(0)}\frac{r_+ j_1(qr)}{r}  
\label{hn}
\end{eqnarray}
\end{widetext}
Here $q=|\vec{q}|$ is the exchange momentum carried by the neutrino propagator. The $g_\alpha(q^2)$'s are the form factors. It can be written in an empiric dipole form in general: $g_V(q^2)=g_V(0) / (1+q^2/\Lambda_V)^2$ and $g_A(q^2)=g_A(0)/(1+q^2/\Lambda_A)^2$. Here $g_V(0)=1$ and $g_A(0)=1.27$, and we take $\Lambda_V=0.85$GeV and $\Lambda_A=1.1$GeV for the energy cutoff. In nuclear environment, $g_A(0)$ is usually quenched with a not definitely known origin. Therefore in the current work we adopt two values for this coupling constant: the bare one and a quenched one with a quenching factor $g_A=0.75 g_{A0}$. And $g_M(q^2)=(1+\kappa_1)g_V(q^2)$\cite{SPV99} with $\kappa_1=\mu_p-\mu_n$. Also $g_P(q^2)=2m_p g_A(q^2)/(q^2+m_\pi^2)$ is given by the PCAC hypothesis.

And the angular operators $\mathcal{O}$s have the forms:
\begin{eqnarray}
\mathcal{O}_1 &=& \vec{\sigma}_1\cdot \vec{\sigma}_2 [\hat{r}\otimes \hat{r}]^{(2)} \nonumber\\
\mathcal{O}_2 &=& [\vec{\sigma}_1 \otimes \vec{\sigma}_2]^{(2)} \nonumber \\
\mathcal{O}_3 &=& [ [\vec{\sigma}_1 \otimes \vec{\sigma}_2]^{(2)}\otimes [\hat{r}\otimes \hat{r}]^{(2)} ]^{(2)} \nonumber \\
\mathcal{O}'_3 &=& [ [\vec{\sigma}_1 \otimes \vec{\sigma}_2]^{(2)}\otimes [ [\hat{r}\otimes \hat{r}]^{(2)}\otimes [\hat{r}\otimes \hat{r}]^{(2)}]^{(4)} ]^{(2)} \nonumber \\
\mathcal{O}_{4} &=& [\hat{r}\otimes \hat{r}]^{(2)} \nonumber \\
\mathcal{O}_{5} &=& [(\vec{\sigma}_1 + \vec{\sigma}_2)\otimes [\hat{r}\otimes \hat{r}]^{(2)}]^{(2)} 
\end{eqnarray}
for the non-primed NMEs $M_\lambda$ and $M_\eta$. 

And
\begin{eqnarray}
\mathcal{O}_{6} &=& [(\vec{\sigma}_1 - \vec{\sigma}_2)\otimes [\hat{r}\otimes \hat{r}_{+}]^{(1)}]^{(2)} \nonumber \\
\mathcal{O}_{7} &=& [(\vec{\sigma}_1 - \vec{\sigma}_2)\otimes [\hat{r}\otimes \hat{r}_{+}]^{(2)}]^{(2)}
\end{eqnarray}
For the primed NME $M_\eta'$.

Compared to the expression in \cite{Tom88}, $M_3$ has one extra term induced by the hadronic current, we denote it by $M_3'$. For $M_4\sim M_7$, no corrections from the induced hadronic current are presented.

For the sake of comparison with the decay to the ground state cases, we find that we have also AA, AP, PP and MM components for the space-space current-current interactions ($M_1$, $M_2$ and $M_3$) coming from different components of induced hadronic current. Except for $M_1$, one finds that PP and MM components are suppressed by a factor smaller than $1/5$ compared to AA and AP components. 

The NME are calculated in our case with the nuclear many-body approach, the so-called pn-QRPA as well as charge conserving QRPA methods both with realistic nuclear forces \ \cite{FF21}. The detailed expression can be found in \ \cite{FF21}:
\begin{eqnarray}
M_{i}&=&\sum_{pnp'n'}^{J^\pi m} \langle 2^+_f|| \widetilde{[c_{p}^\dagger \tilde{c}_{n}]}_{J'} || J^\pi m_f\rangle \langle J^\pi m_f||J^\pi m_f\rangle \nonumber \\
&\times& \langle J^\pi m_i||[c_{p'}^\dagger \tilde{c}_{n'}]_J || 0^+_i \rangle
\end{eqnarray}
The expressions for one body densities as well as the overlap of the initial and final intermediate states can also be found in \ \cite{FF20}, where the final $2^+$ states are obtained by charge conserving QRPA \ \cite{SC93,FF20}.

\begin{table}[htp]
\begin{center}
\caption{The decomposition coefficients of NMEs for $M_\lambda$, $M_{\eta}$ and $M_{\eta}'$. Here $g_A=g_A(0)$.}
\label{coef}
\begin{tabular}{c|ccccccc}
\hline
i & 1 &2 &3 &4 &5 &6 & 7\\
\hline
$C_{\Lambda i}$  & $\frac{1}{3}g_A^2 $ & $-\frac{2}{3}g_A^2$ & $\sqrt{\frac{7}{3}}g_A^2$ & $1$ & $-\sqrt{\frac{3}{2}} g_A$ \\
$C_{\eta i}$ &$\frac{1}{3} g_A^2 $ & $-\frac{2}{3} g_A^2$ & $\sqrt{\frac{7}{3}}g_A^2$ & $-1$ & $0$ \\
$C_{\eta i}'$ & & & & & & $\sqrt{\frac{1}{2}} g_A $ & $-\sqrt{\frac{3}{2}}g_A $ \\
\hline
\end{tabular}
\end{center}
\end{table}

\section{Results and discussions}
\subsection{Phase space factors}
\begin{table*}[htp]
\begin{center}
\caption{PSFs for various nuclei from current work and that of \cite{DKT85} as well as current experimental limits. For results from \cite{DKT85}, we have converted to our convention. We also present the constraints on new physics parameter on two special cases.}
\label{psf}
\begin{tabular}{c|c|cccc|c|cccc}
\hline
& Q(MeV)& \multicolumn{2}{c}{$G_1$ ($10^{-15}y^{-1}$)}& \multicolumn{2}{c|}{$G_2$ ($10^{-15}y^{-1}$) }& $t_{1/2}^{\rm limit}$(yr) & \multicolumn{2}{c}{$|\lambda|\gg |\eta|, m_{\beta\beta}$} & \multicolumn{2}{c}{$|\eta|\gg|\lambda|, m_{\beta\beta}$}\\
&		&	this work & \cite{DKT85} &this work & \cite{DKT85} &  & $2^+$ &$0^+$\cite{SDS15} & $2^+$ & $0^+$\cite{SDS15} \\
\hline
$^{76}$Ge  & 1.480 & 6.86 & 7.37  & 4.77 & 5.12 & $>$2.1$\times$10$^{24}$\cite{MAJ20} & $<$2.13$\times 10^{-5}$ &$<$5.07$\times 10^{-7}$&$<$4.56$\times 10^{-6}$& $<$2.81$\times 10^{-9}$ \\
$^{82}$Se   & 2.219& 50.12 &55.28 & 40.11 & 44.30 & $>$1.0$\times$ 10$^{22}$\cite{Bar17}& $<$8.35$\times 10^{-5}$ &&$<$5.63$\times 10^{-5}$ \\
$^{96}$Zr    & 2.572& 140.4  && 117.2 & & $>$9.1$\times$10$^{20}$\cite{Bar17} & $<$2.76$\times 10^{-3}$ &&$<$2.38$\times 10^{-4}$\\
$^{100}$Mo &2.495 & 134.2  &149.8& 111.3 & 124.3 & $>$1.6$\times$10$^{23}$\cite{Bar17} & $<$2.52$\times 10^{-4}$ &&$<$1.73$\times 10^{-5}$\\
$^{116}$Cd  & 1.520 & 18.18  && 12.98 && $>$6.2$\times$10$^{22}$\cite{Bar17} & $<$1.75$\times 10^{-4}$ &&$<$2.32$\times 10^{-5}$\\
$^{128}$Te  & 0.423&  0.225 & 0.267 & 0.0825 & 0.0985 &  &  & \\
$^{130}$Te  & 1.991 & 76.41 & 91.37 & 60.04 & 71.84 & $>$1.4$\times$10$^{23}$\cite{Bar17} & $<$2.35$\times 10^{-5}$ &&$<$5.11$\times 10^{-6}$\\
$^{136}$Xe  & 1.639 & 35.61 &44.99 & 26.33 & 33.42  & $>$2.6$\times$10$^{25}$\cite{Bar17} & $<$2.43$\times 10^{-6}$ &$<$4.35$\times 10^{-7}$&$<$4.17$\times 10^{-6}$ & $<$2.12$\times 10^{-9}$\\
\hline
\end{tabular}
\end{center}
\end{table*}

For the calculation of PSFs, we use the numerical package {\it Radial} \cite{SFW95} for the electron wave-functions and we follow the convention in\ \cite{KI12}. We use a uniform charge distribution for the calculations of the nuclear static charge potential and we choose the charge radius to be  the same as the nuclear radius.  We neglect the screening effect from the orbital electrons since it gives minor corrections to PSF \ \cite{KI12} in the case of the decay to the ground state. 

The results of PSFs of various nuclei are presented in Table\ \ref{psf}, we also list these nuclei's Q values. For most nuclei except $^{128}$Te, PSFs span a range for about two orders of magnitude. Three nuclei have Q values larger than $2$MeV, and of which $^{96}$Zr has the largest PSF for the decay into the excited state.  $^{100}$Mo has almost the same values for the PSF as $^{96}$Zr, since their Q values are close. While $^{82}$Se has a larger Q value, its PSFs are somehow smaller than that of $^{130}$Te mostly due to its smaller atomic number $Z$.

We also present results from an earlier calculation \ \cite{DKT85}, where a Taylor expansion of the electron wave function is used. For $^{76}$Ge, we have also results from \ \cite{Tom88}, $G_1= 7.34 \times 10^{-15}y^{-1}$ and $G_2= 5.10 \times 10^{-15}y^{-1}$, which are basically the same to results in \cite{DKT85} with a deviation less than 1\% since they use similar treatment for electron wave functions. Our current numerical results generally agree with their results quantitatively. We find the deviations for ours and theirs are about 10\%$\sim$30\%, the current numerical results are smaller than their predictions. There is a strong trend of an increase for the deviations as $Z$ increases. This is reasonable, with the growth of Z, the nuclear radius R also increases. Since the conventional Taylor expansion method uses $\alpha Z$ and $WR$ as variables, the errors will grow with the increase of these variables. One finds the largest phase space factor for $^{100}$Mo from their calculations. If this nucleus has also the largest NME, then it can be one of the most promising candidate for a future experimental search. To explore such possibility, we need high precision nuclear many-body calculations. We will proceed into this direction in the next part.

\subsection{Nuclear matrix elements}

\begin{table*}[htp]
\begin{center}
\caption{NMEs for decay to $2^+_1$ with $g_A=g_{A0}=1.27$, here $a$ and $b$ refer to two different cases with AV-18 and CD-Bonn short range correlations adopted. }
\label{resu}
\begin{tabular}{|cc|cc|cc|cc|cc|cc|cc|cc|cc|}
\hline
          	&	&\multicolumn{2}{c|}{$^{76}$Ge} & \multicolumn{2}{c|}{ $^{82}$Se}  &  \multicolumn{2}{c|}{$^{96}$Zr}   	& \multicolumn{2}{c|}{$^{100}$Mo} 	& \multicolumn{2}{c|}{$^{116}$Cd}  	& \multicolumn{2}{c|}{$^{128}$Te}  	& \multicolumn{2}{c|}{$^{130}$Te}  	& \multicolumn{2}{c|}{$^{136}$Xe}  \\
		&    &	a&b					&	a	&	b	&	a	&	b	&	a	&	b	&	a	&	b	&	a	&	b	& 	a	&	b	&	a	&	b	\\
\hline
 \multirow{5}*{$M_1$}   &AA &0.641 & 0.640 &0.790 & 0.789 &0.027 & 0.027 &0.205 & 0.205 &0.198 & 0.199 &0.700 & 0.700 &0.643 & 0.643 &0.334 & 0.333  \\ 
 &AP &0.481 & 0.480 &0.649 & 0.648 &0.046 & 0.046 &0.141 & 0.141 &-0.058 & -0.057 &0.425 & 0.425 &0.449 & 0.448 &0.563 & 0.562  \\ 
 &PP &-0.087 & -0.087 &-0.149 & -0.149 &-0.022 & -0.022 &-0.022 & -0.022 &0.067 & 0.067 &-0.044 & -0.043 &-0.070 & -0.070 &-0.206 & -0.206  \\ 
 &MM &-0.033 & -0.033 &-0.098 & -0.098 &-0.027 & -0.028 &-0.007 & -0.007 &0.071 & 0.072 &0.018 & 0.018 &-0.016 & -0.016 &-0.182 & -0.183  \\ 
 &tot. &1.001 & 1.000 &1.193 & 1.191 &0.024 & 0.023 &0.317 & 0.317 &0.279 & 0.280 &1.099 & 1.100 &1.006 & 1.006 &0.510 & 0.506  \\ 
\hline 
\multirow{5}*{$M_2$}   &AA&-0.249 & -0.252 &-0.005 & -0.006 &0.238 & 0.239 &-0.052 & -0.054 &-0.992 & -0.996 &-0.268 & -0.269 &-0.175 & -0.175 &0.405 & 0.408  \\ 
 & AP&0.207 & 0.210 &-0.028 & -0.027 &-0.229 & -0.230 &0.265 & 0.267 &0.564 & 0.568 &0.142 & 0.143 &0.059 & 0.060 &-0.416 & -0.419  \\ 
 & PP &-0.034 & -0.034 &-0.002 & -0.003 &0.029 & 0.030 &-0.043 & -0.043 &-0.066 & -0.067 &-0.022 & -0.022 &-0.012 & -0.012 &0.044 & 0.045  \\ 
 & MM &-0.021 & -0.023 &-0.005 & -0.006 &0.014 & 0.014 &-0.022 & -0.023 &-0.035 & -0.037 &-0.014 & -0.015 &-0.008 & -0.009 &0.022 & 0.023  \\ 
 & tot. &-0.097 & -0.098 &-0.041 & -0.042 &0.052 & 0.052 &0.148 & 0.147 &-0.530 & -0.532 &-0.162 & -0.163 &-0.136 & -0.136 &0.056 & 0.058  \\ 
\hline 
\multirow{5}*{$M_3$}   &AA&-0.049 & -0.049 &0.118 & 0.118 &-0.008 & -0.008 &0.158 & 0.158 &0.015 & 0.015 &-0.098 & -0.098 &-0.009 & -0.009 &0.407 & 0.407  \\ 
 &AP &-0.031 & -0.031 &-0.059 & -0.059 &-0.005 & -0.005 &-0.034 & -0.034 &-0.053 & -0.053 &-0.039 & -0.039 &-0.044 & -0.044 &-0.056 & -0.056  \\ 
 & PP &0.004 & 0.004 &-0.000 & -0.000 &-0.001 & -0.001 &-0.001 & -0.001 &0.001 & 0.001 &0.006 & 0.006 &0.003 & 0.003 &-0.009 & -0.009  \\ 
 & MM &0.001 & 0.001 &-0.003 & -0.003 &-0.001 & -0.001 &-0.002 & -0.002 &-0.002 & -0.002 &-0.001 & -0.001 &-0.001 & -0.001 &-0.005 & -0.005  \\ 
 & tot. &-0.074 & -0.074 &0.056 & 0.056 &-0.015 & -0.015 &0.121 & 0.121 &-0.040 & -0.040 &-0.132 & -0.132 &-0.052 & -0.052 &0.336 & 0.336  \\ 
\hline 
\multirow{3}*{$M'_3$}   &PP &-0.003 & -0.003 &-0.023 & -0.023 &0.004 & 0.004 &-0.021 & -0.021 &-0.012 & -0.012 &-0.019 & -0.019 &-0.019 & -0.019 &-0.019 & -0.019  \\ 
 & MM &-0.001 & -0.001 &-0.012 & -0.012 &0.001 & 0.001 &-0.010 & -0.010 &-0.007 & -0.007 &-0.010 & -0.010 &-0.010 & -0.010 &-0.009 & -0.009  \\ 
 & tot. &-0.004 & -0.004 &-0.035 & -0.035 &0.005 & 0.005 &-0.031 & -0.031 &-0.019 & -0.019 &-0.029 & -0.029 &-0.030 & -0.030 &-0.028 & -0.028  \\ 
\hline 
\multicolumn{2}{|c|}{$M_4$}  &-0.147 & -0.147 &-0.123 & -0.123 &-0.072 & -0.071 &0.121 & 0.120 &0.142 & 0.142 &-0.188 & -0.188 &-0.139 & -0.139 &0.119 & 0.119  \\ 
\hline 
 \multicolumn{2}{|c|}{$M_5$} &-0.061 & -0.061 &-0.216 & -0.216 &-0.257 & -0.257 &0.045 & 0.045 &-0.124 & -0.124 &-0.130 & -0.130 &-0.161 & -0.161 &-0.280 & -0.280  \\ 
\hline 
\multicolumn{2}{|c|}{ $M_\lambda$} &0.398 & 0.399 &0.952 & 0.952 &0.262 & 0.261 &0.284 & 0.285 &0.909 & 0.911 &0.384 & 0.385 &0.597 & 0.597 &1.527 & 1.523 \\ 
\multicolumn{2}{|c|}{$M_\eta$}  &0.597 & 0.598 &0.862 & 0.862 &0.005 & 0.004 &0.112 & 0.114 &0.431 & 0.435 &0.558 & 0.559 &0.626 & 0.626 &0.854 & 0.849 \\ 
\hline 
\hline
\multicolumn{2}{|c|}{$M_6$}  &0.631 & 0.640 &0.314 & 0.315 &0.128 & 0.128 &0.481 & 0.484 &-0.609 & -0.615 &0.803 & 0.816 &0.669 & 0.679 &0.054 & 0.049  \\ 
\hline 
\multicolumn{2}{|c|}{$M_7$}  &-1.241 & -1.251 &-0.336 & -0.339 &0.419 & 0.422 &-0.080 & -0.083 &0.948 & 0.951 &-1.752 & -1.762 &-1.346 & -1.353 &0.187 & 0.190  \\ 
\hline 
\multicolumn{2}{|c|}{$M'_\eta$} &2.496 & 2.520 &0.804 & 0.810 &-0.537 & -0.542 &0.557 & 0.564 &-2.021 & -2.032 &3.446 & 3.474 &2.694 & 2.715 &-0.242 & -0.251 \\ 
\hline  
\hline
\end{tabular}
\end{center}
\end{table*}

\begin{table*}[htp]
\begin{center}
\caption{The same table as Table.\ref{resu} but with $g_A=0.75 g_{A0}$ }
\label{resq}
\begin{tabular}{|cc|cc|cc|cc|cc|cc|cc|cc|cc|}
\hline
          	&	&\multicolumn{2}{c|}{$^{76}$Ge} & \multicolumn{2}{c|}{ $^{82}$Se}  &  \multicolumn{2}{c|}{$^{96}$Zr}   	& \multicolumn{2}{c|}{$^{100}$Mo} 	& \multicolumn{2}{c|}{$^{116}$Cd}  	& \multicolumn{2}{c|}{$^{128}$Te}  	& \multicolumn{2}{c|}{$^{130}$Te}  	& \multicolumn{2}{c|}{$^{136}$Xe}  \\
		&    &	a&b					&	a	&	b	&	a	&	b	&	a	&	b	&	a	&	b	&	a	&	b	& 	a	&	b	&	a	&	b	\\
		\hline
 \multirow{5}*{$M_1$}   &AA &0.634 & 0.634 &0.786 & 0.785 &0.027 & 0.027 &0.193 & 0.193 &0.200 & 0.200 &0.699 & 0.698 &0.644 & 0.644 &0.333 & 0.332  \\ 
 &AP &0.476 & 0.476 &0.648 & 0.647 &0.045 & 0.045 &0.137 & 0.137 &-0.058 & -0.057 &0.424 & 0.424 &0.448 & 0.447 &0.563 & 0.561  \\ 
 &PP &-0.086 & -0.086 &-0.149 & -0.149 &-0.022 & -0.022 &-0.022 & -0.022 &0.067 & 0.067 &-0.043 & -0.043 &-0.070 & -0.070 &-0.206 & -0.205  \\ 
 &MM &-0.033 & -0.033 &-0.098 & -0.098 &-0.027 & -0.028 &-0.007 & -0.006 &0.071 & 0.072 &0.018 & 0.019 &-0.016 & -0.015 &-0.181 & -0.183  \\ 
 &tot.&0.991 & 0.991 &1.187 & 1.185 &0.022 & 0.022 &0.302 & 0.302 &0.280 & 0.281 &1.097 & 1.098 &1.007 & 1.006 &0.508 & 0.505  \\ 
\hline 
 \multirow{5}*{$M_2$}   &AA &-0.237 & -0.240 &-0.005 & -0.006 &0.240 & 0.241 &-0.119 & -0.121 &-0.962 & -0.966 &-0.272 & -0.273 &-0.172 & -0.173 &0.404 & 0.407  \\ 
 &AP &0.204 & 0.207 &-0.027 & -0.025 &-0.226 & -0.227 &0.277 & 0.279 &0.553 & 0.558 &0.144 & 0.145 &0.059 & 0.060 &-0.414 & -0.418  \\ 
 &PP &-0.033 & -0.034 &-0.003 & -0.003 &0.029 & 0.029 &-0.043 & -0.044 &-0.066 & -0.066 &-0.022 & -0.023 &-0.012 & -0.012 &0.044 & 0.045  \\ 
 &MM &-0.021 & -0.022 &-0.005 & -0.006 &0.013 & 0.014 &-0.022 & -0.023 &-0.035 & -0.037 &-0.014 & -0.015 &-0.008 & -0.009 &0.022 & 0.023  \\ 
 &tot. &-0.087 & -0.089 &-0.039 & -0.040 &0.056 & 0.057 &0.092 & 0.091 &-0.509 & -0.511 &-0.165 & -0.165 &-0.134 & -0.134 &0.056 & 0.058  \\ 
\hline 
 \multirow{5}*{$M_3$}   &AA &-0.049 & -0.049 &0.118 & 0.118 &-0.007 & -0.007 &0.149 & 0.149 &0.018 & 0.018 &-0.099 & -0.100 &-0.010 & -0.010 &0.407 & 0.406  \\ 
 &AP &-0.031 & -0.031 &-0.059 & -0.058 &-0.005 & -0.005 &-0.034 & -0.034 &-0.054 & -0.054 &-0.039 & -0.038 &-0.044 & -0.044 &-0.056 & -0.056  \\ 
 &PP &0.004 & 0.004 &-0.000 & -0.000 &-0.001 & -0.001 &-0.001 & -0.001 &0.001 & 0.001 &0.006 & 0.006 &0.003 & 0.003 &-0.009 & -0.009  \\ 
 &MM &0.001 & 0.001 &-0.003 & -0.003 &-0.001 & -0.001 &-0.002 & -0.002 &-0.002 & -0.002 &-0.001 & -0.001 &-0.001 & -0.001 &-0.005 & -0.005  \\ 
 &tot. &-0.074 & -0.074 &0.056 & 0.056 &-0.014 & -0.014 &0.112 & 0.112 &-0.038 & -0.038 &-0.133 & -0.133 &-0.053 & -0.053 &0.336 & 0.336  \\ 
\hline 
 \multirow{3}*{$M'_3$}   &PP &-0.003 & -0.003 &-0.023 & -0.023 &0.004 & 0.004 &-0.021 & -0.021 &-0.013 & -0.012 &-0.019 & -0.019 &-0.019 & -0.019 &-0.019 & -0.019  \\ 
 &MM &-0.001 & -0.001 &-0.012 & -0.012 &0.001 & 0.001 &-0.010 & -0.010 &-0.007 & -0.007 &-0.010 & -0.010 &-0.010 & -0.010 &-0.009 & -0.009  \\ 
 &tot. &-0.004 & -0.004 &-0.035 & -0.035 &0.005 & 0.005 &-0.031 & -0.031 &-0.020 & -0.020 &-0.029 & -0.029 &-0.030 & -0.030 &-0.028 & -0.028  \\ 
\hline 
\multicolumn{2}{|c|}{$M_4$} &-0.146 & -0.146 &-0.123 & -0.122 &-0.071 & -0.071 &0.120 & 0.120 &0.143 & 0.143 &-0.187 & -0.188 &-0.140 & -0.140 &0.119 & 0.119  \\ 
\hline 
\multicolumn{2}{|c|}{$M_5$}&-0.061 & -0.062 &-0.214 & -0.214 &-0.258 & -0.257 &0.048 & 0.048 &-0.124 & -0.124 &-0.128 & -0.128 &-0.161 & -0.161 &-0.280 & -0.279  \\ 
\hline 
\multicolumn{2}{|c|}{$M_\lambda$} &0.161 & 0.162 &0.489 & 0.489 &0.194 & 0.193 &0.170 & 0.171 &0.568 & 0.569 &0.133 & 0.133 &0.280 & 0.280 &0.943 & 0.941 \\ 
\multicolumn{2}{|c|}{$M_\eta$} &0.391 & 0.392 &0.535 & 0.535 &0.032 & 0.031 &0.027 & 0.028 &0.170 & 0.172 &0.395 & 0.395 &0.411 & 0.411 &0.427 & 0.425 \\ 
\hline 
\hline
 \multicolumn{2}{|c|}{$M_6$} &0.632 & 0.641 &0.312 & 0.312 &0.128 & 0.127 &0.479 & 0.482 &-0.612 & -0.618 &0.799 & 0.813 &0.671 & 0.681 &0.052 & 0.047  \\ 
\hline 
\multicolumn{2}{|c|}{$M_7$} &-1.242 & -1.252 &-0.333 & -0.336 &0.422 & 0.424 &-0.094 & -0.097 &0.948 & 0.952 &-1.748 & -1.758 &-1.347 & -1.355 &0.187 & 0.190  \\ 
\hline 
\multicolumn{2}{|c|}{$M'_\eta$} &1.874 & 1.893 &0.598 & 0.602 &-0.406 & -0.410 &0.432 & 0.437 &-1.518 & -1.526 &2.577 & 2.598 &2.024 & 2.040 &-0.183 & -0.190 \\ 
\hline 
\hline 
\end{tabular}
\end{center}
\end{table*}

While PSF can be fairly well determined with decent accuracy within about several percents, the NME's have larger uncertainties, greater than a factor of two \cite{YMN21}. In current work we adopt QRPA for the many-body calculations. Compared to the LSSM approach, QRPA has the advantage of small computation requirement with the price of less accuracy. 

For the QRPA method with realistic forces several renormalization parameters are needed to reproduce experimental data. We fit the pairing parameters by the experimental pairing gaps. And we renormalize the residual forces by $g_{pp}^{T=1}$ to eliminate $M_F^{2\nu}$ and $g_{pp}^{T=0}$ to reproduce the measured $2\nu\beta\beta$ NME.

All our results for the NME's are presented in Table \ \ref{resu} and Table \ \ref{resq}. The two tables correspond to two possible values of the axial vector coupling constant $g_A$ commonly used in the  literature. In these tables we also give the results with two different src's (short range correlations) marked as $a$ and $b$ for AV-18 and CD-Bonn\ \cite{SFR09}, respectively and we find these different src's barely change the results. As we are aware, different choices of src's will change the NME's for $0\nu\beta\beta(0^+)$ by about $5-10\%$. Therefore, $0\nu\beta\beta(2^+)$ is less sensitive to the choice of of the src's. This suggests that the two nucleons involved in this decay are more far away from each other, as we shall see in a quantitive analysis later.


As shown in eq. \eqref{hn}, except for $M_1$, the PP and MM terms are suppressed by a factor of about $1/5\sim 1/10$. Therefore, since all the terms related to spherical Bessel function $j_{3}$ in $M_3$ are contained in PP and MM terms, they are heavily suppressed. Even not suppressed by this small factor, the MM terms are generally smaller for $M_1$. In the contrary, for most cases, the AP term plays an important role and they are supposed to be the LO contribution \cite{CDV17}. 

We first discuss the bare $g_A$ case. For $M_1$, besides the LO AA contribution, the major correction of about 2/3 comes from AP term as mentioned above. For most nuclei, this term leads to an enhancement for the final results but for $^{116}$Cd, it gives an 30\% reduction. For $^{136}$Xe, it gives the largest contribution even larger than AA, this makes it the dominant term in $M_1$ for this nucleus. Meanwhile for most nuclei, the PP term gives the reductions from 10\%$\sim$20\% relative to AA term. Two exceptions are the $^{116}$Cd and $^{136}$Xe cases: for the former, this term gives an enhancement about 30\% relative to the AA term which nearly cancels the reduction from the AP term as we discussed above; for the latter, PP gives a 2/3 reduction relative to AA, this term and the AP term together gives nearly the same contribution as AA. The MM terms are supposed to be NLO contributions \ \cite{CDV17}. For $M_1$, it gives negligible contributions for the nuclei $^{100}$Mo, $^{128}$Te and $^{130}$Te, while corrections for the other nuclei ranges from 5\% to 50\%. For the extreme case of $^{96}$Zr, the MM term has an equal size as AA due to the smallness of later. 

For $^{96}$Zr, $M_1$ is heavily suppressed, it is one order of magnitude smaller than other nuclei, and a careful check suggests that this smallness comes from the cancellation from different intermediate states, this differs from the case of $^{76}$Ge, where different intermediate states add up together (see fig.1 of \cite{FF21} ). For all other nuclei, the total $M_1$ is generally within one order of magnitude, {\it e.g.} for $^{76}$Ge, $^{82}$Se, $^{128}$Te and $^{130}$Te, its value is around $1$.


The case for $M_2$ is quite different from $M_1$. For the AA component, different nuclei differ by several orders of magnitude, especially for $^{82}$Se, it is heavily suppressed by the cancellations from different intermediate states. This cancellation also reduces the NME for $^{100}$Mo. For other nuclei it has a value of about $0.2 \sim 0.3$, but for the two cases with semi magic nuclei involved in the transition process, the results are somehow enhanced. Further investigation is needed for the possible relation between the magicity and this enhancement.

The correction from the pseudo-scalar current is much more pronounced for $M_2$. For $M_1$ the AP term is about one order of magnitude larger than the PP term. Unlike for $M_1$, AP terms here mostly appear for  cancellations, this causes an overall smallness of $M_2$ for most nuclei compared to $M_1$. The semi magic nuclei  - as already  mentioned above - have generally large AA terms, however for $^{136}$Xe, the AP term exactly cancels the AA term. This leads to a suppressed $M_2$; meanwhile for $^{116}$Cd, the cancellation from AP is smaller compared to AA. Thus we have the largest $M_2$ for this nucleus. Like for $^{136}$Xe, a nearly exact cancellation between AA and AP terms also happens for $^{76}$Ge and $^{96}$Zr. For $^{100}$Mo, we find that the contribution from AP is about 5 times larger than that from AA. While for Te isotopes, AP reduces the results by about 50\%. The PP term behaves like NLO contributions by the suppression from a small coefficient, their contributions are generally within  a magnitude of 10\%. So does the NLO MM term which contributes with about 10\% for most cases.

Thus for most nuclei, $M_2$ is around 0.1 due to the cancellations between AA and AP terms, with the exception of $^{82}$Se where all the components are small. $^{116}$Cd is the one with the largest $M_2$ about $0.5$. As a result, our calculations suggest that $M_2$ is generally one order of magnitude smaller than $M_1$.

For most nuclei, $M_3$ is generally smaller than $M_1$ but similar to $M_2$. However, unlike the smallness caused by the cancellation between AA and AP terms for $M_2$, the magnitude of each component of $M_3$ is generally much smaller than the counterparts in $M_1$ or $M_2$. For $^{100}$Mo and $^{136}$Xe, the situation is a bit different: AA terms for them are close to that of $M_1$, but on the other hand, there is no large enhancement from AP term for $M_3$, therefore their $M_3$ are generally smaller. And we find for $M_3$, PP and MM terms are negligible. In general, except $^{136}$Xe, $M_3$'s are around or smaller than $0.1$.

We also find that the induced $M'_3$ is much smaller and they barely give visible contributions. But for $^{82}$Se and $^{130}$Te, $M'_3$ is comparable to $M_3$, and gives cancellation for the former nucleus and enhancement for the latter.

In general, for the space-space components of hadronic currents, the inclusion of pseudo-scalar and weak-magnetism terms especially the AP term will change the NME drastically and these changes are usually not negligible. Whether these contributions are positive or negative is nucleus dependent.

$M_4$ comes from the time-time component of hadronic currents and the induced current will not contribute to this component. As in \cite{FF21}, they are generally with the magnitude of $0.1$ and the difference for different nuclei are generally within a factor of two. The largest value is found for  $^{128}$Te and the smallest from $^{96}$Zr, differed nearly by a factor of 3.

$M_5$ is the space-time component of hadronic currents. Their values are generally close to $M_4$ with an magnitude around 1, but differ in details, such as the phases. The relative magnitude of these two NME's is nucleus dependent.

$M_\lambda$ and $M_\eta$ are induced from the coupling of $\vec{q}$ and the relative coordinate of the two decaying nucleons $\vec{r}$. The difference of $M_\lambda$ and $M_\eta$ comes from the space-time and time-time components of the nuclear currents, it can be defined as $\delta M=2 M_{4} - \sqrt{3/2} g_A M_{5}$. This suggests that for nuclei with $M_4$ and $M_{5}$ close to each other, such as $^{82}$Se or $^{130}$Te, this difference is small. On the other hand, this difference is also related to $g_A$. For all nuclei in our calculation, $M_\lambda$ and $M_\eta$ are with the same phase, this means they will cancel or add up with each other depending on the relative sign of $\langle \eta\rangle $ and $\langle \lambda \rangle$. For $^{76}$Ge, $^{82}$Se, $^{128}$Te and $^{130}$Te, $M_\lambda$ is smaller than $M_\eta$, while for other nuclei, $M_\lambda$ is larger. In general, except for $^{136}$Xe, $M_\lambda$ and $M_\eta$ is about one order of magnitude smaller than $M_{0\nu}(0^+)$ (To compare our results with those in various literatures, we need to divide current results by $g_{A0}^2$). $^{136}$Xe does have $M_{\lambda}$ larger than 1 in our convention, but it is still less than half of the values for $M_{0\nu}(0^+)$.

$M'_\eta$ are actually induced by the coupling of the $\vec{q}$ term in the neutrino propagator and the COM (Center Of Mass) coordinate of the two decaying nucleons $\vec{r}_{+}$. The differences of $M_6$ and $M_7$ originate from the coupled angular momenta of the COM and relative coordinate, thus are closely related to the orbitals of the decaying nucleons. In our calculation, we find for most nuclei, these two terms give coherent contribution but for $^{96}$Zr and $^{136}$Xe, they cancel with each other leading to suppressed $M'_{\eta}$'s. For most nuclei, we have also larger $M_7$, but for $^{96}$Zr, $M_7$ is suppressed and is much smaller than $M_6$. Also in most cases, $M'_\eta$ is much larger than $M_\lambda$ and $M_\eta$. Instead of one order of magnitude smaller than corresponding $M^{0\nu}$($0^+$), they are generally smaller within a factor of one half. This differs with previous studies \ \cite{Tom88} where these NMEs are supposed to be several order of magnitude smaller than $M^{0\nu}(0^+)$ and hence can be safely neglected. We will study the consequence of this dominance for the role of determining new physics parameters in the following section.

As presented in Table \ \ref{resu}, different src's produce actually negligibly smaller deviations like for the light neutrino mass mechanism of $0\nu\beta\beta(0^+)$. On the other hand, a comparison of Table \ \ref{resu} and Table \ \ref{resq} suggests that quenching of $g_A$ affects the final results drastically. In QRPA calculations the quenching affects the final results two-fold: \\
First we fit our model parameter $g_{pp}^{T=0}$ by the $2\nu\beta\beta$ NME which is related to $g_A$. Therefore, quenching of $g_A$ will affect the choice of this parameter and subsequently the individual NMEs. \\
Second  the quenching changes the coefficients $C$'s in Table. \ref{coef} which are than multiplied with the individual partial NME's to get the final total NME's. \\
For the current calculation, we find that the change of the parameter $g_{pp}^{T=0}$ induced by quenching will not largely change the individual NMEs, their changes are within 10\%. Hence the change of the final results can be directly connected to the quenched value of $g_A$ in the coefficients $C$.

For $M_{\lambda}$, only the coefficient of $M_4$ are independent of $g_A$, and all other coefficients have a dependence linear or squared on $g_A$. Therefore, the quenching generally reduce $M_\lambda$ by a magnitude from 30\% to 50\% for different nuclei. The largest reduction in percentage is found for $^{128}$Te, where the NME for the quenched case is only one third of that of the bare $g_A$ case. The reduction for $M_\eta$ is similar to that of $M_{\lambda}$, we now have no $g_A$ linear dependent terms any more. The general reduction is from 30\% to 70\%, the largest reduction is found for  $^{100}$Mo with more than 70\%. While for $^{96}$Zr an enhancement is observed, but for this nucleus, all the individual NME's are small and they cancel each other significantly, the presence of quenching largely reduce the cancellation of $M_1\sim M_3$ to $M_4$ which then leads to the enhancement. All these behaviors for various nuclei stem from the interplay between $M_4$ and other individual NME's which are reduced by $g_A$. Drastic reductions are also found for $M'_\eta$ with a nearly fixed factor $q=g_A/g_{A0}$ since the two terms are both proportional to $g_A$ and the $g_{pp}^{T=0}$ dependence is weak for $M_6$ and $M_7$. Therefore, as for the decay to the ground states, the role of $g_A$ is important. We need to understand the origin of the quenching and the precise value of $g_A$ for the $\beta\beta$-decay, we also need to determine wether the quenching is operator dependent.

\subsection{Constraints on L-R symmetric model}
\begin{figure}
\includegraphics[scale=0.55]{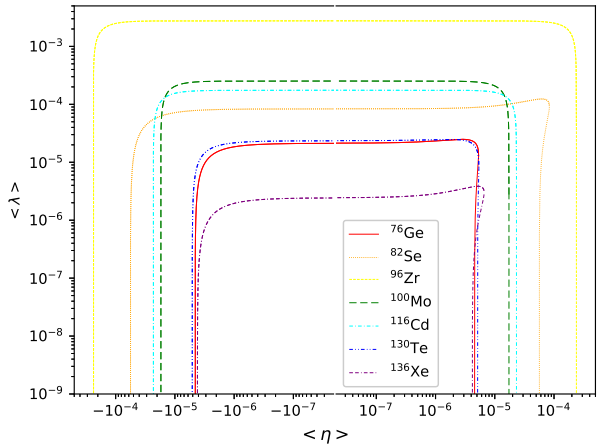}
\caption{(Color online) Constraints on $\langle \lambda \rangle$ and $\langle \eta \rangle$ from $0\nu\beta\beta(2^+)$-decay. {The region outside the curves are excluded by the current limits obtained from the measurements and our calculations.}}
\label{cstr}
\end{figure}

Unlike for the decay to ground state, where the neutrino mass mechanism is supposed to be dominant, the neutrino mass mechanism for $0\nu\beta\beta(2^+)$ can only be triggered by nuclear recoil \ \cite{Tom99} and hence is suppressed by a factor of $Q/2M$ compared to $\langle \lambda \rangle$ and $\langle \eta \rangle$ terms. In \ \cite{Tom99}, an estimation suggests that their NMEs are about several orders of magnitude smaller than that of LR mechanism. Since this NME for the mass mechanism in the decay to excited state is extremely hindered, we do not expect to to be able to observe this decay mode with the dominance of the mass mechanism.

Therefore measurements of $0\nu\beta\beta(2^+)$ can be used as perfect constraints on new physics parameters such as $\langle \lambda \rangle$ or $\langle \eta \rangle$. In this section, with the calculated NME and PSF as well as the experimental limit presented in Table.\ref{psf}, we perform a simple analysis on several extreme cases. Currently the most stringent constrain for half-lives is that of $^{136}$Xe, which pushes the limit to $10^{25}$y while others are two or three orders of magnitude shorter. For $^{96}$Zr, the lower limit is $10^{20}$ years, which is much shorter than all others. {While \cite{SDS15} provides a way of probing different mechanism from comparisons of different nuclei, current method would help distinguish the underlying new physics within one isotope. This is done by comparing the decay rates to the ground and excited state as we shall show.}

\subsubsection{$|\lambda| \gg |\eta|$}
In the $\lambda$ dominant case (See Table.\ref{psf}), {eq.\eqref{dw} becomes $\Gamma= G_1 M_\lambda^2 |\langle \lambda \rangle|^2$ and}  the most stringent constraints for $\lambda$ are obtained from $^{136}$Xe. With our calculated NME's, we obtain the constraint $|\lambda|<2.43\times 10^{-5}$. Other nuclei yield  generally the same magnitude except $^{96}$Zr and $^{100}$Mo which sets limits one order of magnitude larger. Since all these nuclei except $^{136}$Xe have basically similar $M_\lambda$, these larger limits generally are due to shorter half-life limits.  In Table \ \ref{psf}, we present constraints obtained in \cite{SDS15} from decay to ground states for two nuclei from QRPA calculations without considering the hadronic induced currents. With a later calculation \ \cite{SSD17}, one finds that the NME for $0\nu\beta\beta(0^+)$ is slightly enhanced, but this will not change the general magnitude of the constraints. Therefore, we can make a direct comparison. Our results suggest that for the $\lambda$ dominant case, the constraints for $^{76}$Ge differs by two orders of magnitude while for $^{136}$Xe this difference is less than one order of magnitude while comparing our results with those from \ \cite{SDS15}. The half-life lower limits used for $0\nu\beta\beta(0^+)$ for both nuclei are around $10^{25}$ yr, while for $0\nu\beta\beta(2^+)$, $^{76}$Ge has a half-life lower-limit one order of magnitude shorter. However, $^{136}$Xe has the similar half-life lower limits to the $2^+$ and to the ground state. Combining these results, we find that for certain nuclei ($^{136}$Xe here), for the case of $\lambda$ dominance, the half-lives for the two modes are within a difference of two orders of magnitude, this could perhaps be used in the future for the identification of the decay mechanism if both modes are observed.

\subsubsection{$|\lambda|\ll |\eta|$}
For this case, {eq.\eqref{dw} becomes $\Gamma=(G_1 M_\eta^2+G_2 M'^2_\eta) |\langle \eta \rangle|^2$ and} the most stringent constraints come also from $^{136}$Xe. However, $^{76}$Ge and $^{130}$Te yield  also constraints close to $^{136}$Xe. They all require that $|\eta|$ is smaller than about  $10^{-6}$. However, these constraints are far looser than that of $0\nu\beta\beta(0^+)$, where a limit around $|\eta|<10^{-9}$ is obtained assuming the dominance of the $\eta$ mechanism. If this case is true in nature, it will not be possible to observe the $2^+$ decay in these nuclei, {since we have an suppression for about 6 orders of magnitude for $0\nu\beta\beta(2^+)$ compared to $0\nu\beta\beta(0^+)$ for $^{136}$Xe and 4$\sim$5 orders of magnitude for other nuclei. If future measurements do push the half-life limit of $0\nu\beta\beta(2^+)$ to 4$\sim$5 orders of magnitude longer, we can then rule out both $\lambda$ and $\eta$ mechanisms.}

\subsubsection{$|\lambda|\sim |\eta|$}
More general cases are presented in fig.\ref{cstr}. {For these cases, since $\langle \eta \rangle$ and $\langle \lambda \rangle$ share the same phase $e^{i\psi}$, eq.\eqref{dw} can be written explicitly as $\Gamma=G_1 M_\lambda^2  |\langle \lambda \rangle|^2+(G_1 M_\eta^2+G_2 M_\eta^2) |\langle \eta \rangle|^2 - 2 G_1  M_\lambda M_\eta \langle \lambda \rangle \langle \eta \rangle$. In the current convention, $\lambda$ is always positive and the sign for $\eta$ is not definite. Since different nuclei give constraints of $\lambda$ and $\eta$ on different orders of magnitude, we use a logarithm scale in the graph. The interference term of $\eta$ and $\lambda$ rotates the long axis of the ellipse to the first quadrant.} From the figure, we can clearly see that, like in the $\lambda$ or $\eta$ dominant cases, $^{136}$Xe sets the most stringent constraints. While $^{76}$Ge and $^{130}$Te set both less tight constraints for $\lambda$ and $\eta$. And it is obvious that the future slight improvement of the measurement will surely set more stringent constraints for these two nuclei. This is highly probable for $^{130}$Te, of which current half-life lower limit is about two orders of magnitude smaller than that for $^{136}$Xe. And better half-life lower limits for other nuclei will also improve our analysis.  A combined analysis with decay to ground states could give us more hints on the relative magnitude of $\langle \eta \rangle$ and $\langle \lambda \rangle$.

{In general, the measurement of decay to the $2^+$ excited states offers a way for the discrimination of underlying mechanism. For example, for $^{136}$Xe, within our calculations, a  half-time lower limit for about 2 orders of magnitude longer than the observed half-life of $0\nu\beta\beta(0^+)$ will rule out the possibility of the existence of the $\lambda$ mechanism dominance. However, these conclusions rely heavily on the NME calculations, and need further verifications from other many-body calculations. }
 
\section{Conclusions and Outlooks}
In this work, we systematically investigate the decay rates of $0\nu\beta\beta(2^+)$ under the LR symmetric model, where both the $\lambda$ and $\eta$ mechanisms are involved. {We incorporate the contributions from induced currents and find that the pseudo-scalar current is important for NME calculations.} Our results suggest that, the NME is nucleus dependent and may differ by up to one order of magnitude, also the NME for the two different mechanisms in the same nucleus can be differed by more than one order of magnitude. {We also improve the phase space factor calculations by using numerical electron wave functions.} These results lead to different constraints on different nuclei both from different NME's and different current lower limits of decay half-lives. With the comparison to $0\nu\beta\beta(0^+)$, we can constrain the underlying mechanisms within individual nucleus. These comparisons suggest that for the special $\lambda$ dominant case, $^{136}$Xe has the potential of comparable decay half-lives for decays to both ground state and $2^+$ excited state. Future experiments could shift the lower limit up and set more stringent constraints on the new physics parameters for other nucleus, and even have the potential of discovering the possible existence of right-handed gauge bosons. {Meanwhile, to draw a more solid conclusion, more nuclear many-body calculations are needed to make a better prediction of the NMEs which are keys for these investigations.}

\section*{acknowledgement}
This work is supported by the "Light of West China" and "From Zero to One" programs from CAS and Guangdong Major Project of Basic as well as Applied Basic Research No. 2020B0301030008 from Guangdong Province. The numerical calculations in this paper have been carried out on the supercomputing system in the Southern Nuclear Science Computing Center. This work is also supported by the National Key Research and Development Program of China (2021YFA1601300).

\end{document}